\documentclass{article}
\usepackage{spconf,amsmath,amssymb,math,graphicx,enumitem}
 
\title{Semi-supervised Sound Event Detection using Random Augmentation and Consistency Regularization}
\name{Xiaofei Li}
\address{School of Engineering, Westlake University, Hangzhou, China \\
         	Institute of Advanced Technology, Westlake Institute for Advanced Study, Hangzhou, China}
\begin{document}
%
\maketitle
\begin{abstract}
Sound event detection is a core module for acoustic environmental analysis. Semi-supervised learning technique allows to largely scale up the dataset without increasing the annotation budget, and recently attracts lots of research attention. In this work, we study on two advanced semi-supervised learning techniques for sound event detection. Data augmentation is important for the success of recent deep learning systems. This work studies the audio-signal random  augmentation method, which provides an augmentation strategy that can handle a large number of different audio transformations. In addition, consistency regularization is widely adopted in recent state-of-the-art semi-supervised learning methods, which exploits the unlabelled data by   constraining the prediction of different transformations of one sample to be identical to the prediction of this sample. This work finds that, for semi-supervised sound event detection, consistency regularization is an effective strategy, especially the best performance is achieved when it is combined with the MeanTeacher model.     
\end{abstract}
\vspace{-0.2cm}
\section{Introduction}
\label{sec:intro}
\vspace{-0.2cm}

Sound event detection (SED) temporally locates and recognizes the sound event from an audio stream, which plays a critical role in automatic analysis of acoustic environments  \cite{virtanen2018}. In recent years, deep neural network has became the dominant technique for SED, since its powerful data representation capability naturally match with the high complexity/diversity of acoustic data. In \cite{piczak2015,salamon2017}, convolutional neural network (CNN) was applied on the audio spectrogram to perform sound classification, which treats spectrogram as an image. Sound classification predicts the class label of audio clips, but not the temporal location of the event, which is thus referred to as weak-prediction. \cite{su2017} proposed to perform sound event temporal detection, namely providing frame-level strong-prediction, using only weakly-labelled (clip-level annotated) data. DCASE (Detection and Classification of Acoustic Scenes and Events) 2017 Challenge task 4 \cite{mesaros2017} released a similar task with \cite{su2017}, namely SED with weakly-labelled data. To largely scale up the dataset without increasing the annotation budget, a large amount of unlabelled data were involved for training since DCASE 2018 \cite{serizel2018}, which arose the problem of semi-supervised learning, namely only a portion of the training data are annotated. The winning system of DCASE 2018, i.e. \cite{jiakai2018}, used a convolutional-recurrent neural network (CRNN) to model both the local spectra and the temporal dynamic of audio signal. To perform semi-supervised learning, \cite{jiakai2018} adopted the MeanTeacher network \cite{tarvainen2017}. This architecture, i.e. CRNN plus MeanTeacher, and its variants are adopted in the baseline and top-performance systems of DCASE 2019 task 4 \cite{turpault2019,delphin2019}, and of DCASE 2020 task 4 \cite{miyazaki2020}. The major improvements over this architecture include applying data augmentation \cite{delphin2019} and using more powerful network, such as Transformer \cite{miyazaki2020}. 

Semi-supervised learning recently attracts lots of attention in the deep learning community \cite{xie2019,berthelot2019,sohn2020}. Semi-supervised learning needs to provide an artificial label for unlabelled data. Pseudo-label \cite{lee2013} takes the argmax of the current network prediction as the artificial label, which transforms the most probable prediction as a hard label. This principle is also shared by entropy minimization \cite{grandvalet2005} and label sharpening used in \cite{xie2019,berthelot2019}. Another important technique generating artificial label is to build a teacher network based on the being-trained (student) network, such as by ensembling \cite{laine2016} or exponentially smoothing (MeanTeacher) \cite{tarvainen2017} the networks of previous training steps. The prediction of teacher network is taken as the artificial label of the student network. Data augmentation largely improves the data variability and constantly improves the system performance, which is widely adopted in recent developed semi-supervised methods \cite{piczak2015,salamon2017,su2017,delphin2019,miyazaki2020,xie2019,berthelot2019,sohn2020}.     AutoAugment \cite{cubuk2019} provides an automated augmentation strategy with reinforcement learning. RandAugment \cite{cubuk2019} is one recent proposed effective and easy-to-use data augmentation mechanism. It randomly selects one transformation for each sample at each training step, which allows to exploit a large number of different types of transformations without increasing the training complexity. Consistency regularization \cite{sajjadi2016} constrains the prediction of different transformations of one sample to be identical to the prediction of this sample, which regularizes the network parameters to be more robust to data disturbance, and becomes an important component in recent studies \cite{berthelot2019,sohn2020,sajjadi2016,hu2017}. 

The augmentation of image data has been intensively investigated for various computer vision tasks \cite{cubuk2019}. In the previous SED methods \cite{piczak2015,salamon2017,su2017,delphin2019,miyazaki2020}, different audio transformations have been tested, and the scale of each transformation is empirically set. In this work, RandAugment is studied for audio data augmentation, which considers a large number of widely used audio transformations, including signal speeding, time shifting \cite{delphin2019}, time stretching \cite{mcfee2015}, pitch shifting \cite{mcfee2015}, dynamic range compression (DRC) \cite{mcfee2015}, time/frequency masking \cite{park2019} and mixup \cite{zhang2017mixup}. MeanTeacher is currently the most popular semi-supervised learning mechanism for SED. In this work, following the research trend of the state-of-the-art semi-supervised learning, consistency regularization is studied. It is found that consistency regularization performs well for audio data, it solely already outperforms MeanTeacher, and can further improve the performance when combined with MeanTeacher.   

\vspace{-0.1cm}
\section{Method}
\label{sec:format}
\vspace{-0.2cm}

In this work, the multi-class sound event detection problem is considered, which means multiple events could be concurrent. In the time-frequency domain, let $\mathbf{x}_t|_{t\in[1,T]} \in \mathbb{R}^{K\times 1}$ denote the feature vector of one utterance, where $T$ and $K$ denote the number of time frames and frequencies, respectively. Three types of data are used: i) strongly-labelled data $\{\mathbf{x}^{(s)}_{n,t}|_{t\in[1,T]} \in \mathbb{R}^{K\times 1}, y^{(s)}_{n,t,c}|_{t\in[1,T'],c\in[1,C]} \in \{0,1\}\}_{n=1}^{N_s}$, where $N_s$ and $C$ denote the data number of one batch and the number of classes, respectively. The strong label $\mathbf{y}^{(s)}_{n,t,c}$ is given for each time frame $t$. Note that the number of output frames $T'$ could be smaller than the one of input feature, i.e. $T$, to have a coarser time resolution;  ii) weakly-labelled data $\{\mathbf{x}^{(w)}_{n,t}|_{t\in[1,T]} \in \mathbb{R}^{K\times 1}, y^{(w)}_{n,c}|_{c\in[1,C]} \in \{0,1\} \}_{n=1}^{N_w}$, where only the weak label $y^{(w)}_{n,c}$ is given for the entire utterance; and iii) unlabelled data  $\{\mathbf{x}^{(u)}_{n,t}|_{t\in[1,T]} \in \mathbb{R}^{K\times 1}\}_{n=1}^{N_u}$, where no labels available. 

\vspace{-0.2cm}
\subsection{Semi-supervised Training Loss}
\vspace{-0.2cm}
For any one labelled or unlabelled utterance $\mathbf{x}_{n}$, we let $\hat{y}^{(s)}_{n,t,c}(\mathbf{x}_{n}) \in [0,1]$ and $\hat{y}^{(w)}_{n,c}(\mathbf{x}_{n}) \in [0,1]$ denote the network prediction of strong labels and weak labels for this utterance, respectively. The supervised classification loss considering both strongly-labelled and weakly-labelled data is 
\begin{align}
\label{eq:super}
\mathcal{L}_{\text{super}}=&\frac{1}{N_sT'C}\sum_{n=1}^{N_s}\sum_{t=1}^{T'}\sum_{c=1}^{C}H(y^{(s)}_{n,t,c},\ \hat{y}^{(s)}_{n,t,c}(\mathbf{x}^{(s)}_{n})) \nonumber \\
&+ \frac{1}{N_wC}\sum_{n=1}^{N_w}\sum_{c=1}^{C}H(y^{(w)}_{n,c}, \ \hat{y}^{(w)}_{n,c}(\mathbf{x}^{(w)}_{n})),
\end{align}
where $H(\cdot)$ denotes binary cross-entropy. 

To exploit the unlabelled data, MeanTeacher model \cite{tarvainen2017} is used to provide pseudo labels. In practice, the pseudo labels will be applied not only to the unlabelled data, but also to the labelled data to improve the training stability. Let $\tilde{y}^{(s)}_{n,t,c}(\mathbf{x}_{n}) \in [0,1]$ and $\tilde{y}^{(w)}_{n,c}(\mathbf{x}_{n}) \in [0,1]$ denote the MeanTeacher strong and weak pseudo-labels, respectively. The unsupervised mean squared error is then: 
\begin{align}
\label{eq:unsuper}
\mathcal{L}_{\text{unsuper}}=&\frac{1}{NT'C}\sum_{n=1}^{N}\sum_{t=1}^{T'}\sum_{c=1}^{C}(\tilde{y}^{(s)}_{n,t,c}(\mathbf{x}_{n})- \hat{y}^{(s)}_{n,t,c}(\mathbf{x}_{n}))^2 \nonumber \\
&+ \frac{1}{NC}\sum_{n=1}^{N}\sum_{c=1}^{C}(\tilde{y}^{(w)}_{n,c}(\mathbf{x}_{n})- \hat{y}^{(w)}_{n,c}(\mathbf{x}_{n}))^2,
\end{align}
where $N=N_s+N_w+N_u$. Data augmentation could largely increase the data diversity and thus improve the performance. Augmented data are generated by applying signal transformations on the original data, and inherit the labels of the original data. The supervised and unsupervised losses defined in (\ref{eq:super}) and (\ref{eq:unsuper}) can be directly applied to the augmented data. 

The augmented data have to be identified as the same class with the corresponding original data, which is implemented by consistency regularization. In practice, it is found that better performance can be achieved when consistency regularization is applied to both the labelled and unlabelled data, and to both the strong and weak predictions. For utterance $\mathbf{x}_{n}$, let $\alpha_p(\mathbf{x}_{n}), p\in[1,P]$ denote its $P$ different transformations. The consistency regularization term is defined as: 
\begin{align}
\label{eq:cr}
&\mathcal{L}_{\text{cr}}=  \nonumber \\
&\frac{1}{NPT'C}\sum_{n=1}^{N}\sum_{p=1}^{P}\sum_{t=1}^{T'}\sum_{c=1}^{C}L(\hat{y}^{(s)}_{n,t,c}(\mathbf{x}_{n})-\hat{y}^{(s)}_{n,t,c}(\alpha_p(\mathbf{x}_{n})))^2 \nonumber \\
&+ \frac{1}{NPC}\sum_{n=1}^{N}\sum_{p=1}^{P}\sum_{c=1}^{C}L(\hat{y}^{(w)}_{n,c}(\mathbf{x}_{n})-\hat{y}^{(w)}_{n,c}(\alpha_p(\mathbf{x}_{n})))^2.
\end{align}
Finally, the overall loss is set to 
\begin{align}
\mathcal{L} = \mathcal{L}_{\text{super}} + \lambda_{\text{unsuper}}\mathcal{L}_{\text{unsuper}} +\lambda_{\text{cr}}\mathcal{L}_{\text{cr}},
\end{align}
where $\lambda_{\text{unsuper}}$ and $\lambda_{\text{cr}}$ are predefined weights.

\subsection{Random Data Augmentation}
\label{sec:rda}

This work follows the RandAugment \cite{cubuk2020} principle. The $P$ transformations are randomly selected from a total of $Q$ available transformations with a uniform distribution, thus there are $Q^P$ potential policies for one utterance. This random-selection is independently performed for each utterance at each training epoch. A number of widely used data transformations are tested. A proper distortion magnitude should be chosen for each transformation. Searching the optimal magnitude for each individual transformation has a very large search space. In RandAugment \cite{cubuk2020}, it was proposed to use a single global distortion magnitude for all the transformations, which largely reduce the search space. This work sets 10 integer distortion scales for each transformation, and the optimal global scale is set by grid-searching from 1 to 10. The audio transformation schemes used in this work include:
\begin{itemize}[leftmargin=*,itemsep=0pt]
\item Signal speeding slows down or speeds up the original signal, which were conducted by up-sampling or down-sampling the signal. This transformation changes the signal length, and also shifts the frequencies. Ten up-sampling factors are set from 1.05 to 1.5 with 0.05 increment. In addition, the factor is randomly set as its reciprocal with 0.5 probability to account for the down-sampling case. 
\item Time shifting rolls the signal along time \cite{delphin2019,koh2020}. The rolling factor is randomly selected from 0.1 to 0.9. The distortion scale is fixed.
\item Time stretching raises or lowers the speed and keeps the original pitch. The audio degradation toolkit \cite{mcfee2015} is used to conduct this transformation. The stretching factors is set as the same with the sampling factors of signal speeding. 
\item Pitch shifting \cite{mcfee2015} raises or lowers the pitch and keeps the original signal length. Ten positive shifting scales are set from 0.5 to 5 with 0.5 increment. The corresponding negative factors are randomly used with 0.5 probability. 
\item Dynamic range compression (DRC) \cite{mcfee2015}. One mode is randomly chosen for each utterance. The distortion scale is fixed.
\item Time masking is a spectral augmentation technique proposed in \cite{park2019} for speech recognition, which masks a period of consecutive time frames to 0. In this work, a masking unit is set as a period of consecutive time frames with the duration of 0.05 times the total signal length. The masking scales are set as taking randomly positioned 1 to 10 masking units.   
\item Frequency masking \cite{park2019} masks frequencies to 0. The masking scales are set following the spirit of time masking.
\item Mixup \cite{zhang2017mixup} takes the convex combination of two samples (and corresponding labels) as a new sample (and label). In this work,  the sum of two utterances is considered as concurrent sound events, and the two mixed utterances are not rescaled. The mixup label is set by taking the \emph{logical alternation} of the two original labels. The mixup sample is generated using two samples from the same training mini-batch. In the consistency regularization loss (\ref{eq:cr}), the predictions $\hat{y}^{(s)}_{n,t,c}(\mathbf{x}_{n})$ and $\hat{y}^{(w)}_{n,c}(\mathbf{x}_{n})$ actually should be the mixup of the prediction of the two mixed samples, which is computed by first binarizing the prediction of the two mixed samples and then taking the \emph{logical alternation}. The distortion scale is fixed for mixup.   
\end{itemize} 

\section{Experiments}
\label{sec:exp}

In this work, we use the dataset of DCASE 2020 task 4 "Sound event detection and separation in domestic environments" \cite{dcase2020}, which includes 10 domestic sound classes: speech, dog, cat, alarm/bell/ringing, dishes, frying, blender, running water, vacuum cleaner, electric shaver/toothbrush. The training dataset (we have downloaded) consists of weakly-labelled data of 1466 clips, synthetic strongly-labelled data of 2584 clips and unlabelled in domain data of 13343 clips. The validation set includes 1168 clips of real-recorded signals with strong annotations. Out of the synthetic 2584 clips, 517 clips are used for training validation. The 1168 validation clips are used for test. The length of all these clips are 10 s.

The DCASE 2020 task 4 official baseline system \footnote{https://github.com/turpaultn/dcase20\_task4} is adopted to develop the proposed method, which is a modification of \cite{delphin2019}. The sampling rate is 16 kHz. The 128-dimensional feature is extracted in the short-time Fourier transform (2048 window, 255 hop size) domain with mel-scale frequency bins. The mean-teacher model \cite{jiakai2018} is adopted for semi-supervised learning. The network includes 7 layers of CNNs and two layers of GRU-RNNs \cite{delphin2019}. A median filter with duration of 0.45 second is used for post-processing. The batch size is 24, and each batch is composed of 6 weakly-labelled, 6 strongly-labelled and 12 unlabelled samples. The number of training epochs is set to 200. In this work, the learning rate scheme is set as: rampuping to $10^{-3}$ at epoch 50, step decaying to $2\times10^{-4}$ at epoch 100, and further decaying to $4\times10^{-5}$ at epoch 150. The weights $\lambda_{\text{unsuper}}$ and $\lambda_{\text{cr}}$ are rampupped to a constant at epoch 50, then kept invariant. 

The performance is evaluated with three metrics: i) the macro-averaging event-based collar F1 score \cite{mesaros2016}. A 200 ms collar on onsets and a 200 ms and 20\% of the events length collar on offsets are used for the comparison between event prediction and ground truth; ii) the macro-averaging event-based PSDS (polyphonic sound detection score) F1 score and cross-trigger (CT) PSDS F1 score \cite{bilen2020}. To have a reliable evaluation, each of the following experiments are run three independent trials, and the averaged scores are reported.

\subsection{SED Results} 

The sound event detection results are given in Table \ref{tab:res}. MT and MT+RDA stand for the baseline MeanTeacher method without and with random data augmentation (RDA), respectively, for which $\lambda_{\text{unsuper}}=2$ and $\lambda_{\text{cr}}=0$. CR+RDA stands for consistency regularization excluding MeanTeacher, with $\lambda_{\text{unsuper}}=0$ and $\lambda_{\text{cr}}=2$, as consistency regularization itself is also an unsupervised learning strategy. Finally, MT+CR+RDA combines MeanTeacher and consistency regularization, with $\lambda_{\text{unsuper}}=2$ and $\lambda_{\text{cr}}=2$. We test two network architectures with different activations for CNN layers, i.e. GLU (Gated Linear Units) and CG (Context Gating). It can be seen that the proposed techniques work well for  both  network architectures, and CG consistently performs better than GLU. MT+RDA largely improves the performance of MT, especially for the (CT) PSDs scores, which shows the efficacy of random data augmentation. CR+RDA  outperforms MT+RDA. It means consistency regularization solely is even better than MeanTeacher in the sense of exploiting unlabelled data, which is consistency with the image classification results presented in \cite{sohn2020}. MT+SC+RDA provides the best scores, which indicates that the two unsupervised losses, i.e. MeanTeacher and self-consistency, are somehow complementary. 

Besides, we have also studied several other semi-supervised learning techniques, including hard pseudo-label \cite{sohn2020,lee2013}, entropy minimization \cite{grandvalet2005,miyato2018}, information maximization \cite{hu2017}, and their combination with others. However, we did not find a better strategy than the combination of MeantTeacher and consistency regularization. In the literature, many different combination strategies of semi-supervised learning techniques have been reported, and achieved superior performance on various tasks, especially on the computer vision tasks. However, it seems  that one strategy can hardly keep on top of a wide range of tasks. One needs to carefully investigate the proper strategy for one specific task. 

\begin{table}[t]
\centering
\caption{Sound event detection results.}
\label{tab:res}
\begin{tabular}{c c  | c | c | c  }  
 \multicolumn{2}{c|}{F1 score (\%)}   & collar   & PSDS  & CT PSDS  \\ \hline
 & MT     & 37.2 & 60.8 &  53.5 \\
GLU & MT+RDA  & 39.4  &  64.3   & 57.5  \\  
 & CR+RDA   & 39.3 &   66.2 &   60.0 \\
 & MT+CR+RDA & 40.7  &  66.5 &   60.7 \\  \hline
 & MT  & 38.8  &  61.9 &  55.2 \\
CG & MT+RDA & 40.8  &  66.8 &  61.2 \\
 & CR+RDA   & 41.2 &    67.1 &   61.3 \\
 & MT+CR+RDA   & 43.5 &  69.5 &   64.4 \\  
\end{tabular}
\vspace{-.3cm}
\end{table}

%

\subsection{Setup for Random Data Augmentation}

In this section, the random data augmentation method is studied in more detail. Based on preliminary experiments, the number of transformations applied to each sample, i.e. $P$, is set to 1, which will not be analyzed in detail, due to the room limit. 
The magnitude for the audio transformations listed in Section \ref{sec:rda} should be empirically set, and independently setting for each one leads to a very large search space. In RandAugment, the optimal transformation magnitude is searched with a global scale as defined in Section \ref{sec:rda}. Table \ref{tab:scale} lists the grid search results with the CG MT+CR+RDA method, for the scales of 3, 4, 5 and 6. Two schemes are tested: fixed scale and random scale with a fixed upper bound. It can be seen that random scale averagely outperforms fixed scale. Random scale 5 achieves the best performance, which is thus used in all of other experiments. It was demonstrated in \cite{cubuk2020} that changing the magnitude for one transformation does not largely affect the performance. In addition, each type of the transformation should be assured to play a positive role. This is done by comparing the results using all of them and the results using all excluding each one of them. The results with the GLU CR+RDA method are given in Table \ref{tab:trans}.  It is seen that excluding mixup or pitch shifting largely degrades the performance relative to the 'all' case, which means they are very useful. Excluding DRC or frequency masking achieves similar scores with the 'all' case, thence they are not really functional in this experiment. The other four transformations improve the performance to a certain extent, and thus have a medium importance.  


\begin{table}[t]
\centering
\caption{Global grid-search for random augmentation.}
\label{tab:scale}
\begin{tabular}{c c  | c  c  c  }  
 \multicolumn{2}{c|}{F1 score (\%)}   & collar   & PSDS  & CT PSDS  \\ \hline
 & 3     & 41.9 & 68.4  & 63.0 \\
fixed & 4  & 41.8 & 69.0  & 64.0  \\  
scale & 5   & 41.5 &  68.2  &  62.9 \\
 & 6 & 41.6  &  68.6 &   63.1 \\  \hline
 & 3  & 41.9  &  67.8 &  62.2 \\
random & 4 & 42.1 & 68.5  &  63.2 \\
scale & 5   & \textbf{43.5}  &  \textbf{69.5}  &  \textbf{64.4} \\
 & 6   & 42.5 &  69.3 &   64.1 \\  
\end{tabular}
\vspace{-.3cm}
\end{table}

\begin{table}[t]
\centering
\caption{Results for excluding one transformation. }
\label{tab:trans}
\begin{tabular}{c  | c | c | c  }  
 F1 score (\%)   & collar   & PSDS  & CT PSDS  \\ \hline
all     & 39.3 &  66.2 &   60.0 \\
- Signal speeding  & 38.3 &  66.0  &  60.0  \\  
- Time shifting  & 38.6 &  65.4  &  59.7 \\
- Time stretching & 38.7 & 65.8  &  59.1 \\  
- Pitch shifting  & 37.7 & 65.7 &  59.4 \\
- DRC & 38.8    & 66.5 &    60.3 \\
- Time masking  & 38.4    & 65.5   &  59.7 \\
- Frequency masking  & 39.2 &    66.0  &  60.3 \\  
- Mixup & 37.1 &    64.0 &    57.7 \\
\end{tabular}
\vspace{-.3cm}
\end{table}

\section{CONCLUSIONS}
\label{sec:foot}

This work has studied the random data augmentation strategy with a number of different audio transformations. When proper parameters are chosen, random augmentation noticeably improves the SED performance. For augmented data, consistency regularization is adopted as an effective unsupervised loss. The combination of consistency regularization and MeanTeacher achieves the best performance. Note that this work focuses only on the semi-supervised learning strategies, and many other techniques not adopted in this work may can further improve the performance, such as in the DCASE 2020 winning system \cite{miyazaki2020} that better network, better post-processing median filter or multi-system ensembling are used.     




\small

\bibliographystyle{IEEEbib}
\bibliography{lixf_bib}

\begin{thebibliography}{10}

\bibitem{virtanen2018}
Tuomas Virtanen, Mark~D Plumbley, and Dan Ellis,
\newblock {\em Computational analysis of sound scenes and events},
\newblock Springer, 2018.

\bibitem{piczak2015}
Karol~J Piczak,
\newblock ``Environmental sound classification with convolutional neural
  networks,''
\newblock in {\em MLSP}, 2015, pp. 1--6.

\bibitem{salamon2017}
Justin Salamon and Juan~Pablo Bello,
\newblock ``Deep convolutional neural networks and data augmentation for
  environmental sound classification,''
\newblock {\em IEEE Signal Processing Letters}, vol. 24, no. 3, pp. 279--283,
  2017.

\bibitem{su2017}
Ting-Wei Su, Jen-Yu Liu, and Yi-Hsuan Yang,
\newblock ``Weakly-supervised audio event detection using event-specific
  gaussian filters and fully convolutional networks,''
\newblock in {\em ICASSP}, 2017, pp. 791--795.

\bibitem{mesaros2017}
Annamaria Mesaros, Toni Heittola, Aleksandr Diment, Benjamin Elizalde, Ankit
  Shah, Emmanuel Vincent, Bhiksha Raj, and Tuomas Virtanen,
\newblock ``Dcase 2017 challenge setup: Tasks, datasets and baseline system,''
\newblock in {\em DCASE Challenge}, 2017.

\bibitem{serizel2018}
Romain Serizel, Nicolas Turpault, Hamid Eghbal-Zadeh, and Ankit~Parag Shah,
\newblock ``Large-scale weakly labeled semi-supervised sound event detection in
  domestic environments,''
\newblock {\em arXiv preprint arXiv:1807.10501}, 2018.

\bibitem{jiakai2018}
Lu~JiaKai,
\newblock ``Mean teacher convolution system for dcase 2018 task 4,''
\newblock {\em DCASE Challenge}, 2018.

\bibitem{tarvainen2017}
Antti Tarvainen and Harri Valpola,
\newblock ``Mean teachers are better role models: Weight-averaged consistency
  targets improve semi-supervised deep learning results,''
\newblock in {\em Advances in neural information processing systems}, 2017, pp.
  1195--1204.

\bibitem{turpault2019}
Nicolas Turpault, Romain Serizel, Justin Salamon, and Ankit~Parag Shah,
\newblock ``Sound event detection in domestic environments with weakly labeled
  data and soundscape synthesis,''
\newblock {\em DCASE Challenge}, 2019.

\bibitem{delphin2019}
Lionel Delphin-Poulat and Cyril Plapous,
\newblock ``Mean teacher with data augmentation for dcase 2019 task 4,''
\newblock {\em Orange Labs Lannion, France, Tech. Rep}, 2019.

\bibitem{miyazaki2020}
Koichi Miyazaki, Tatsuya Komatsu, Tomoki Hayashi, Shinji Watanabe, Tomoki Toda,
  and Kazuya Takeda,
\newblock ``Convolution-augmented transformer for semi-supervised sound event
  detection,''
\newblock Tech. {R}ep., DCASE Challenge, 2020.

\bibitem{xie2019}
Qizhe Xie, Zihang Dai, Eduard Hovy, Minh-Thang Luong, and Quoc~V Le,
\newblock ``Unsupervised data augmentation for consistency training,''
\newblock {\em arXiv preprint arXiv:1904.12848}, 2019.

\bibitem{berthelot2019}
David Berthelot, Nicholas Carlini, Ekin~D Cubuk, Alex Kurakin, Kihyuk Sohn, Han
  Zhang, and Colin Raffel,
\newblock ``Remixmatch: Semi-supervised learning with distribution alignment
  and augmentation anchoring,''
\newblock {\em arXiv preprint arXiv:1911.09785}, 2019.

\bibitem{sohn2020}
Kihyuk Sohn, David Berthelot, Chun-Liang Li, Zizhao Zhang, Nicholas Carlini,
  Ekin~D Cubuk, Alex Kurakin, Han Zhang, and Colin Raffel,
\newblock ``Fixmatch: Simplifying semi-supervised learning with consistency and
  confidence,''
\newblock {\em arXiv preprint arXiv:2001.07685}, 2020.

\bibitem{lee2013}
Dong-Hyun Lee,
\newblock ``Pseudo-label: The simple and efficient semi-supervised learning
  method for deep neural networks,''
\newblock in {\em ICML}, 2013, vol.~3.

\bibitem{grandvalet2005}
Yves Grandvalet and Yoshua Bengio,
\newblock ``Semi-supervised learning by entropy minimization,''
\newblock in {\em Advances in neural information processing systems}, 2005, pp.
  529--536.

\bibitem{laine2016}
Samuli Laine and Timo Aila,
\newblock ``Temporal ensembling for semi-supervised learning,''
\newblock in {\em Internation Conference on Learning Representation}, 2017.

\bibitem{cubuk2019}
Ekin~D Cubuk, Barret Zoph, Dandelion Mane, Vijay Vasudevan, and Quoc~V Le,
\newblock ``Autoaugment: Learning augmentation strategies from data,''
\newblock in {\em IEEE conference on computer vision and pattern recognition},
  2019, pp. 113--123.

\bibitem{sajjadi2016}
Mehdi Sajjadi, Mehran Javanmardi, and Tolga Tasdizen,
\newblock ``Regularization with stochastic transformations and perturbations
  for deep semi-supervised learning,''
\newblock in {\em Advances in neural information processing systems}, 2016, pp.
  1163--1171.

\bibitem{hu2017}
Weihua Hu, Takeru Miyato, Seiya Tokui, Eiichi Matsumoto, and Masashi Sugiyama,
\newblock ``Learning discrete representations via information maximizing
  self-augmented training,''
\newblock 2017, vol.~70 of {\em Proceedings of Machine Learning Research}, pp.
  1558--1567.

\bibitem{mcfee2015}
Brian McFee, Eric~J Humphrey, and Juan~Pablo Bello,
\newblock ``A software framework for musical data augmentation.,''
\newblock in {\em ISMIR}, 2015, vol. 2015, pp. 248--254.

\bibitem{park2019}
Daniel~S Park, William Chan, Yu~Zhang, Chung-Cheng Chiu, Barret Zoph, Ekin~D
  Cubuk, and Quoc~V Le,
\newblock ``Specaugment: A simple data augmentation method for automatic speech
  recognition,''
\newblock in {\em Interspeech}, 2019.

\bibitem{zhang2017mixup}
Hongyi Zhang, Moustapha Cisse, Yann~N Dauphin, and David Lopez-Paz,
\newblock ``mixup: Beyond empirical risk minimization,''
\newblock in {\em International Conference on Learing Representations}, 2018.

\bibitem{cubuk2020}
Ekin~D Cubuk, Barret Zoph, Jonathon Shlens, and Quoc~V Le,
\newblock ``Randaugment: Practical automated data augmentation with a reduced
  search space,''
\newblock in {\em Proceedings of the IEEE/CVF Conference on Computer Vision and
  Pattern Recognition Workshops}, 2020, pp. 702--703.

\bibitem{koh2020}
Chih-Yuan Koh, You-Siang Chen, Shang-En Li, Yi-Wen Liu, Jen-Tzung Chien, and
  Mingsian~R Bai,
\newblock ``Sound event detection by consistency training and pseudo-labeling
  with feature-pyramid convolutional recurrent neural networks,''
\newblock {\em DCASE Challenge}, 2020.

\bibitem{dcase2020}
``http://dcase.community/challenge2020/task-sound-event-detection-and-separation-in-domestic-environments,''
\newblock .

\bibitem{mesaros2016}
Annamaria Mesaros, Toni Heittola, and Tuomas Virtanen,
\newblock ``Metrics for polyphonic sound event detection,''
\newblock {\em Applied Sciences}, vol. 6, no. 6, pp. 162, 2016.

\bibitem{bilen2020}
{\c{C}}a{\u{g}}da{\c{s}} Bilen, Giacomo Ferroni, Francesco Tuveri, Juan
  Azcarreta, and Sacha Krstulovi{\'c},
\newblock ``A framework for the robust evaluation of sound event detection,''
\newblock in {\em IEEE International Conference on Acoustics, Speech and Signal
  Processing (ICASSP)}, 2020, pp. 61--65.

\bibitem{miyato2018}
Takeru Miyato, Shin-ichi Maeda, Masanori Koyama, and Shin Ishii,
\newblock ``Virtual adversarial training: a regularization method for
  supervised and semi-supervised learning,''
\newblock {\em IEEE transactions on pattern analysis and machine intelligence},
  vol. 41, no. 8, pp. 1979--1993, 2018.

\end{thebibliography}

\end{document}